\documentclass[conference]{IEEEtran}

\usepackage{cite}
\usepackage{amsmath,amssymb,amsfonts}
\usepackage{graphicx}
\usepackage{textcomp}
\usepackage{xcolor}
\usepackage{booktabs}
\usepackage{multirow}
\usepackage{array}
\usepackage{balance}
\usepackage[breaklinks=true]{hyperref}
\usepackage{url}

\newcommand{\vect}[1]{\boldsymbol{#1}}
\newcommand{\cps}{\ensuremath{\,\mathrm{cycles/sample}}}

\begin{document}

\title{Prediction-Induced Sparsity for Tracked FMCW Radar:
A Track-Conditioned Event-Gap FPGA Architecture}

\author{
\IEEEauthorblockN{Huy Trinh}
\IEEEauthorblockA{
\textit{Department of Electrical and Computer Engineering}\\
\textit{University of Waterloo}\\
Waterloo, Canada\\
Email: h3trinh@uwaterloo.ca
}
}

\maketitle

\begin{abstract}
Tracked FMCW radar already carries a prediction of the next target state, yet the downstream digital hardware is commonly provisioned for a fixed spectral search.
We present prediction-induced sparsity (PIS), a tracked-mode architecture that uses the prior \emph{before} event formation and hardware scheduling. The predicted beat is derotated to an innovation-centred coordinate, while tracker covariance is propagated to a bounded residual support $K$. Sparse level-crossing packets are then executed by an exact event-gap resonator engine with a shared complex processing element and $K$-dependent state/coefficient memory. On a public FMCW dataset, causal prior-only conditioning reduces event records by 33.1\% on average, and the representative fixed-point stream is bit-exact in RTL. And on an Artix-7 implementation, reducing physical support from 129 to 23 states lowers LUT and BRAM use by 65.0\% and 88.1\%, respectively, and reduces measured VCCINT power by 19.9\% at matched throughput. The results show how tracking uncertainty can become a hardware-capacity implication for future adaptive radar accelerators.
\end{abstract}

\begin{IEEEkeywords}
FMCW radar, cognitive radar, event-driven processing, FPGA, sparse hardware, resonator, target tracking
\end{IEEEkeywords}

\section{Introduction and Related Work}
Conventional frequency-modulated continuous-wave (FMCW) radar is usually arranged around acquisition. When the target state is unknown, a broad range--Doppler search is appropriate because the receiver must preserve many possible hypotheses. However, after detection and association, the processing problem changes: a tracker already predicts where the next return should appear and also carries a measure of how uncertain that prediction is. Repeating the same broad processing before using this information can therefore leave both computation and hardware capacity disconnected from the information already available in the radar loop. 

This observation follows the closed-loop principle of cognitive radar. 
Haykin described a cognitive radar as a transmitter--environment--receiver loop in which a Bayesian tracker preserves and updates knowledge of the scene, and the resulting information can influence subsequent sensing~\cite{1593335}.
Layered cognitive architectures later emphasized that prior knowledge can be used at different time scales for adaptive signal processing and resource management. In detail, sufficiently informative prior knowledge can reduce sensing resources instead of repeatedly solving the full acquisition problem~\cite{Brüggenwirth18}. 

In this paper, PIS adopts a deliberately narrower receiver-side interpretation.
It does not modify the transmitted waveform, but feeds the track prediction into the next receive-processing interval so that both the representation and the local hardware support follow what is already known.
This direction also follows our previous low-resolution radar work~\cite{trinh_rasso}. RASSO reshaped an already formed Doppler representation to emphasize difficult near-zero motion~\cite{trinh_rasso}, 
whereas TCRE moved the track prior before spectral formation and estimated only residual range/Doppler error~\cite{trinh2026trackconditionedresidualfrequencyestimation}. PIS converts that residual-coordinate principle into a circuit architecture: the prior mean controls temporal event activity, while prior covariance controls how many residual resonator states are provisioned.
Event-driven radar and neuromorphic hardware provide the complementary implementation context. Time-coded spiking Fourier processing, spiking neural resonators, activity-gated Loihi~2 radar, adaptive-frequency R\&F processing, NeuroRadar, event-driven radar hardware, near-field neuromorphic imaging, and mixed-signal R\&F circuits all show that sparse temporal representations can be useful for radar and sensing hardware~\cite{9743682,Reeb_2025,Reeb_2026,chiavazza2026adaptivefrequencyresonateandfireneuronsspectral,10.1145/3745033,pavlicek,11461855,11562373}.
PIS differs in where the sparsity is created: the event stream and resonator support are conditioned \emph{before the next observation} by the radar track.
The main contributions of our research can be summarized in two points. 
First, we introduce a track-conditioned event-gap resonator architecture in which mean prediction centers the FMCW beat before level crossing and covariance bounds the local resonator bank.
Second, we implement this architecture as fixed-point RTL and close the radar-to-hardware path through measured I/Q, bit-exact simulation, timing-closed Artix-7 implementations, physical replay, and repeated rail-power measurements.
\section{Prediction-Conditioned Radar Representation}
\begin{figure}[h]
    \centering
    \includegraphics[width=\linewidth]{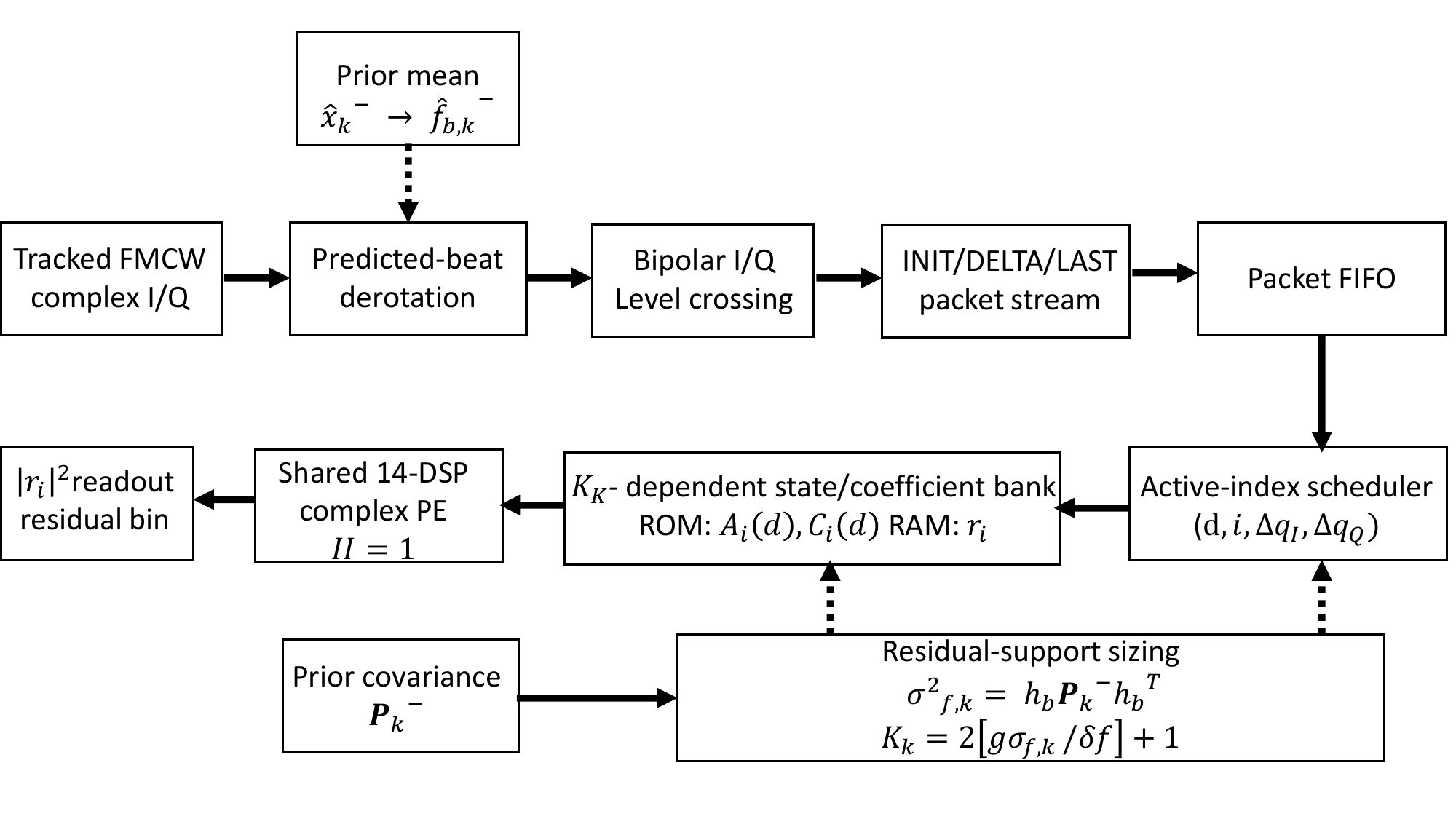}
    \caption{PIS tracked-mode dataflow. The predicted mean centers the next FMCW beat before level-crossing transduction, while the predicted covariance sets the residual support $K$. INIT/DELTA/LAST packets are then executed by a serialized event-gap resonator engine with one shared complex PE and $K$-dependent state/coefficient storage.}
    \label{fig:pis_architecture}
\end{figure}
Figure~\ref{fig:pis_architecture} shows the complete track-mode of PIS. 
The predicted mean centers the next beat before event formation, reducing temporal activity when the prediction is accurate. In parallel, predicted covariance determines a bounded residual-frequency support $K$. The packet scheduler and state/coefficient bank therefore expose two hardware quantities already implicit in the tracker: event traffic and local spectral capacity.
\subsection{Residual coordinate and event activity}
Following the standard single-tone deramped FMCW model~\cite{richards,patole,trinh2026trackconditionedresidualfrequencyestimation}, a tracked component within chirp $k$ can be written as
\begin{equation}
x_k[n]=a_k e^{j(2\pi f_{b,k}nT_s+\phi_k)}+w_k[n],
\label{eq:tone}
\end{equation}
where $f_{b,k}=f_b(\vect x_k)$ is the beat frequency associated with the local target state. After association, the tracker provides a prior state $\hat{\vect x}_k^-$ and therefore a predicted beat $\hat f_{b,k}^-=f_b(\hat{\vect x}_k^-)$. PIS applies the same residual-coordinate idea used in TCRE~\cite{trinh2026trackconditionedresidualfrequencyestimation}, but before event formation:
\begin{equation}
z_k[n]=x_k[n]e^{-j2\pi\hat f_{b,k}^{-}nT_s}
\approx a_ke^{j(2\pi\Delta f_{b,k}nT_s+\phi_k)}+\eta_k[n],
\label{eq:derot}
\end{equation}
where $\Delta f_{b,k}=f_{b,k}-\hat f_{b,k}^{-}$. 
The derotation remove deterministic phase evolution that is already predicted. Therefore, a prior presents a slowly varying innovation signal to the event encoder. For $c\in\{I,Q\}$, let $h_c$ denote the last reconstructed level and $\Theta$ the level-crossing step, a bipolar level crossing with held value $h_c$ and step $\Theta$ generates
\begin{equation}
q_c[n]=\operatorname{sgn}(d_c[n])
\left\lfloor\frac{|d_c[n]|}{\Theta}\right\rfloor,
\qquad d_c[n]=z_c[n]-h_c .
\label{eq:lc}
\end{equation}
A nonzero $q_I$ or $q_Q$ creates one \emph{event record}: its magnitude gives the signed I/Q threshold quanta. This is an event-based representation in the same sense that event/spike interfaces communicate changes
, but the implemented resonator core remains a linear state engine rather than a trained Spiking Neural Network (SNN).

For the RTL path, I/Q samples and the held level are represented in signed Q2.14 fixed-point. The event stream is 
encoded as three packet types.
An \textsc{INIT} packet loads the initial held I/Q value and clears all resonator states. A \textsc{DELTA} packet carries the sample gap $d$ since the preceding event together with signed quanta $(\Delta q_I,\Delta q_Q)$. A final \textsc{LAST} packet advances every active resonator to the common end-of-chirp sample even when no threshold crossing occurs there. As shown in Fig.~\ref{fig:pis_architecture}, the resonator index $i$  is generated by the active-index scheduler.
\subsection{Prior covariance as a hardware-support variable}

The lower branch of Fig.~\ref{fig:pis_architecture} maps tracking uncertainty to local spectral capacity.
Let the prior error be $\tilde{\vect x}_k=\vect x_k-\hat{\vect x}_k^-$ with covariance $\vect P_k^-$ and beat-frequency Jacobian $\vect h_b=\partial f_b/\partial\vect x$ at the predicted state. Linearizing the FMCW beat map about the predicted state gives
\begin{equation}
f_b(\hat{\vect x}_k^-+\tilde{\vect x}_k)
\approx \hat f_{b,k}^-+\vect h_b\tilde{\vect x}_k,
\qquad
\vect h_b=\left.\frac{\partial f_b}{\partial \vect x}\right|_{\hat{\vect x}_k^-}.
\end{equation}
Consequently, the first-order residual-frequency variance is
\begin{equation}
\sigma_{f,k}^2=\vect h_b\vect P_k^-\vect h_b^T,
\qquad
K_k=2\left\lceil\frac{g\sigma_{f,k}}{\delta f}\right\rceil+1 ,
\label{eq:K}
\end{equation}
where $\delta f$ is the resonator spacing and $g$ sets the desired local coverage. 
 We use $g=3.5$ as a conservative prototype setting, corresponding to $1-2Q(3.5)=99.953\%$ 
 nominal coverage for a scalar Gaussian residual.
Thus, Eq.~\eqref{eq:K} specifies the local tracked-mode support, while acquisition or reacquisition remains available when the prior is invalid or the required support is exceeded.
 The TCRE used the same prior to center a local residual estimator and compared it with a same-prior zoomed spectral search~\cite{trinh2026trackconditionedresidualfrequencyestimation}. PIS has a different hardware implication: the residual span is tied to the number of coefficient/state entries in the local resonator bank, not only to the numerical frequency interval evaluated by a fixed processor.

\section{Event-Gap RTL and FPGA Architecture}
\begin{figure*}[t]
\centering
\includegraphics[width=\linewidth,clip,trim=0 6.5cm 0 6cm]{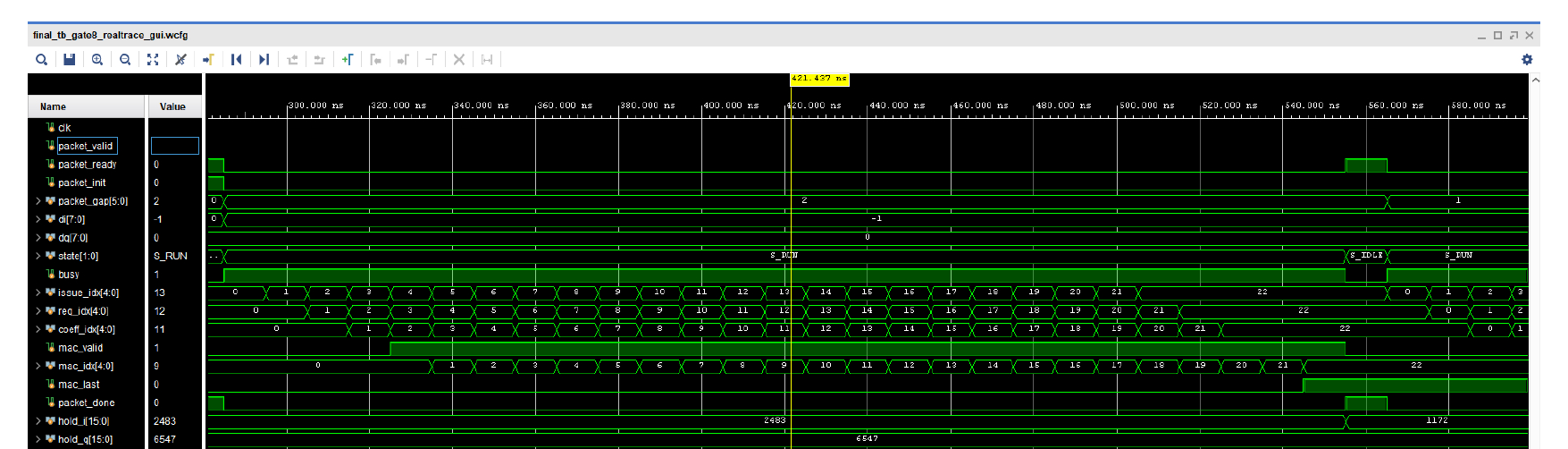}
\caption{Cycle-level RTL execution of a measured-radar \textsc{DELTA} packet with $d=2$ and $(\Delta q_I,\Delta q_Q)=(-1,0)$. State indices 0--22 pass through the request/coefficient/MAC pipeline while the old global hold is preserved. The yellow cursor in XSim inspects at 421.437~ns; after the final MAC/\texttt{packet\_done}, $h_I$ changes from 2483 to 1172 and $h_Q$ remains 6547.}
\label{fig:rtlphys}
\end{figure*}
Once the input has been converted to sparse level-crossing packets, dense zero-valued sample replay 
would ignore the intended event-driven operation.
PIS instead advances each resonator directly across the silent interval. 
We use the linear subthreshold resonant state underlying R\&F spectral processing~\cite{Reeb_2025,chiavazza2026adaptivefrequencyresonateandfireneuronsspectral}. 
For resonator $i$, the held input $h=h_I+jh_Q$ is constant between two events and the corresponding closed-form transition across a gap of $d$ samples is
\begin{equation}
\dot r_i(t)=p_i r_i(t)+h(t),
\qquad
p_i=-\tau^{-1}-j2\pi f_i ,
\end{equation}
where $r_i$ is the complex resonator state. 
Integrating the state equation over $dT_s$ therefore gives the exact piecewise-constant transition
\begin{equation}
\begin{aligned}
r_i^+ &= A_i(d)r_i+C_i(d)h,\\
A_i(d)&=e^{p_i dT_s},\qquad
C_i(d)=\frac{A_i(d)-1}{p_i}.
\end{aligned}
\label{eq:gap}
\end{equation}
The $A_i(d)$ and $C_i(d)$ are gap-dependent transition complex coefficients stored in synchronous ROM and $r_i$ is the complex resonator state stored in RAM.
This formulation preserves the decay and phase accumulated during the gap without reconstructing the missing samples. 
The present design uses the resonant state as a spectral primitive only and it does not implement 
an internal threshold/reset spike generator in each resonator.
The packet ordering is important because the held input is global while the states are independent. 
A \textsc{DELTA} packet applies the same old held input $h$ to all $K$ active states and updates the hold only after the final state commits. With $N_{\rm pkt}$ packets including \textsc{INIT}, the resulting state work is $N_{\rm rec}=(N_{\rm pkt}-1)K$. It is noted that the fixed-point core stores held I/Q in Q2.14, event quanta in signed 8 bits, $A_i(d)$ in 18 bits, $C_i(d)$ in 24 bits, and each complex state in two 32-bit words. One pipelined complex PE processes a resonator state per cycle after fill and maps to 14 DSP48E1 blocks. 

Figure~\ref{fig:rtlphys} show a cycle-level verification of this architecture. 
A measured \textsc{DELTA} packet is serviced across all 23 active states while the global hold remains unchanged, and the hold commits only after \texttt{packet\_done}. 
The yellow cursor in the XSim verifies the scheduling rule required by Eq.~\eqref{eq:gap} and supports the bit-exact closure reported in Table~\ref{tab:meas}.
\begin{table}[h]
\caption{Measured-radar and fixed-point/RTL closure.}
\label{tab:meas}
\centering
\scriptsize
\setlength{\tabcolsep}{3.2pt}
\begin{tabular}{p{0.435\columnwidth}r}
\toprule
Metric & Result\\
\midrule
Post-acquisition mean record reduction & 33.088\%\\
23-state innovation capture & 100\%\\
Fixed-point residual-frequency RMSE & 0.126 FFT bin\\
Representative chirp & 37 (median-like records)\\
Prior / reference frequency & 0.06723923 / 0.06728144 \cps\\
Innovation & $4.2213\times10^{-5}$ \cps\\
Absolute $\rightarrow$ residual records & 227 $\rightarrow$ 154\\
Absolute $\rightarrow$ residual quanta & 654 $\rightarrow$ 208\\
Q2.14 packets / 23-state recurrences & 144 / 3,289\\
XSim final complex states & 23/23 exact\\
Physical replay, P23 and P129/K23 & 100/100 each\\
Returned states / mismatches & 4,600 / 0\\
\bottomrule
\end{tabular}
\end{table}
\section{Measured Radar and Physical FPGA Results}
\begin{figure*}[t]
\centering
\begin{minipage}[t]{0.32\textwidth}\centering
\includegraphics[width=\linewidth]{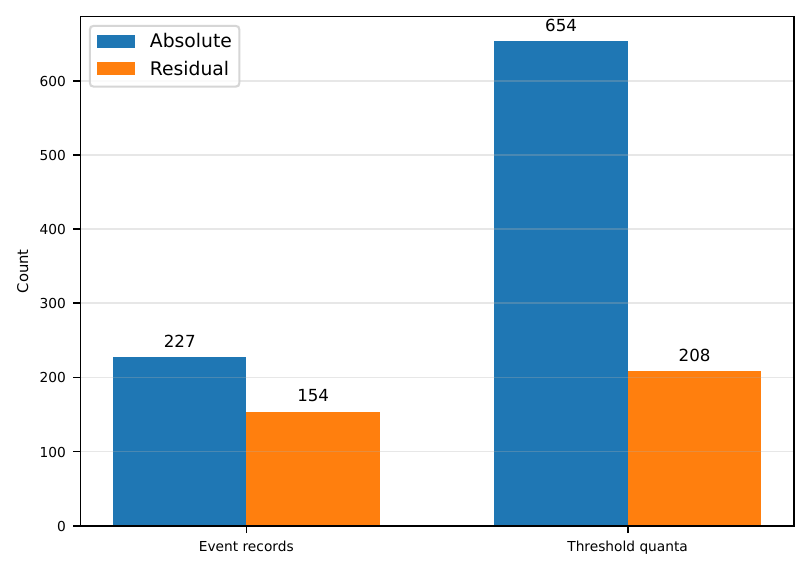}\\[-1mm]
\small (a) measured event workload
\end{minipage}\hfill
\begin{minipage}[t]{0.32\textwidth}\centering
\includegraphics[width=\linewidth]{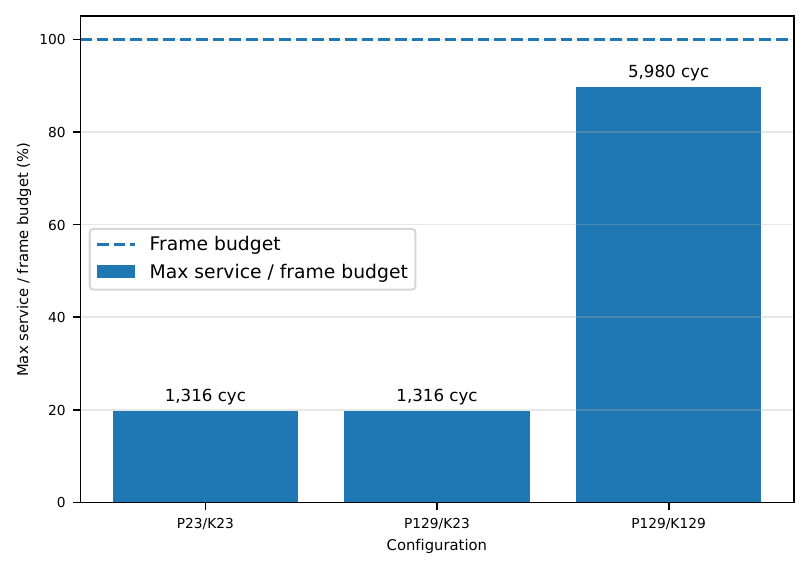}\\[-1mm]
\small (b) physical service margin
\end{minipage}\hfill
\begin{minipage}[t]{0.32\textwidth}\centering
\includegraphics[width=\linewidth]{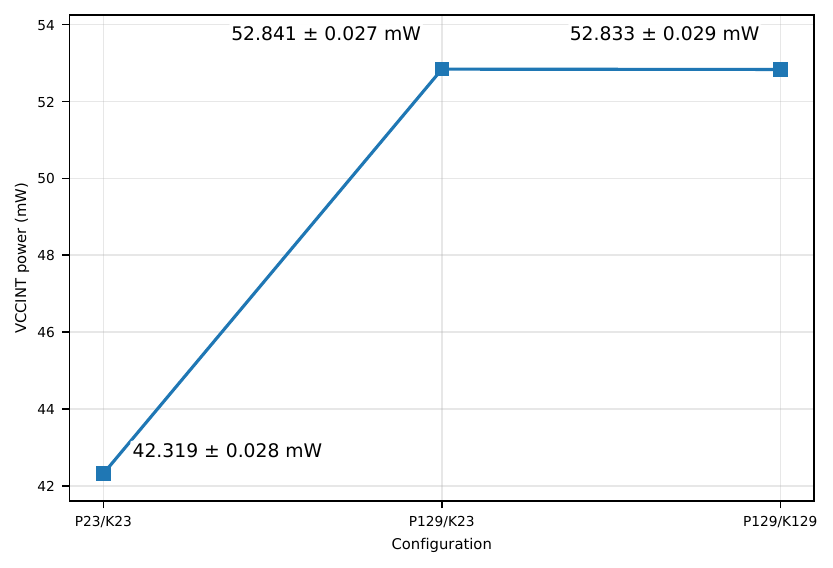}\\[-1mm]
\small (c) matched-rate rail power
\end{minipage}
\caption{Measured-radar to physical-FPGA closure. \textbf{(a)} For representative chirp~37, predicted-beat conditioning changes 227$\rightarrow$154 event records and 654$\rightarrow$208 threshold quanta. \textbf{(b)} P23/K23 and P129/K23 require 1,316 cycles, while P129/K129 requires 5,980 cycles; all remain below the fixed 6,667-cycle frame budget. \textbf{(c)} Five board rounds give 42.319$\pm$0.028, 52.841$\pm$0.027, and 52.833$\pm$0.029~mW, respectively, separating physical right-sizing from runtime narrowing inside the same P129 bitstream.}
\label{fig:results}
\end{figure*}

\subsection{External measured data and bit-exact closure}

We use the public experimental FMCW recording released with the adaptive-frequency R\&F study of Chiavazza \emph{et al.}~\cite{chiavazza2026adaptivefrequencyresonateandfireneuronsspectral,chiavazza2026dataset}. Collected at TU Delft with a TI AWR2243/MMWCAS-RF-EVM platform, it provides an independent complex-I/Q sequence originally released for streaming spectral-estimation research. 
The recording is used only as input while the causal predictor, derotation, level-crossing transduction, fixed-point model, RTL, and FPGA implementation are ours.


One front-end slice contains 128 chirps with 256 fast-time samples. After global 99.5th-percentile normalization, the prior is strictly causal: two-point linear extrapolation is used when available, and the previous estimate otherwise; the current-chirp FFT is used only as the reference/update. Since 
sampling-rate and FMCW-slope metadata were not recovered for this file, errors remain in cycles/sample or FFT bins.

Across post-acquisition chirps, prior conditioning reduces event records by 33.088\% on average while all reference innovations remain inside the 23-state bank. Figure~\ref{fig:results}(a) shows chirp~37, selected because its residual record count is closest to the sequence median. Its 144 packets require 3,289 state recurrences, XSim reproduces all 23 final states exactly, and physical P23/P129-K23 replay returns 4,600 states without mismatch. Table~\ref{tab:meas} summarizes the remaining fixed-point closure.

\subsection{Timing-closed physical capacity and rail power}

We implemented the hardware prototype on a Digilent Arty A7-100T with the Xilinx/AMD Artix-7 XC7A100TCSG324-1 FPGA~\cite{digilent_arty_a7}. And we used Vivado 2025.2 for synthesis, placement, routing, bitstream generation, and hardware access. The core clock is 100~MHz. 
VCCINT is the FPGA core/block-memory rail observed through the board/XADC measurement path~\cite{amd_ug480}. Therefore, it should be interpreted as FPGA rail power only not as total radar-sensor or system power.
\begin{table}[h]
\caption{Timing-closed XC7A100T-1 stand-alone PIS cores.}
\label{tab:resource}
\centering
\scriptsize
\setlength{\tabcolsep}{3.8pt}
\begin{tabular}{lrrr}
\toprule
Metric & P23 & P129 & P23 reduction\\
\midrule
Slice LUTs & 492 & 1,405 & 65.0\%\\
LUT as memory & 132 & 792 & 83.3\%\\
BRAM tiles & 5 & 42 & 88.1\%\\
DSP48E1 & 14 & 14 & 0\%\\
WNS @ 100 MHz & +2.737 ns & +1.921 ns & both pass\\
\bottomrule
\end{tabular}
\end{table}

\begin{figure}[th]
\centering
\includegraphics[width=\columnwidth, clip, trim = 12cm 0.65cm 0 0]{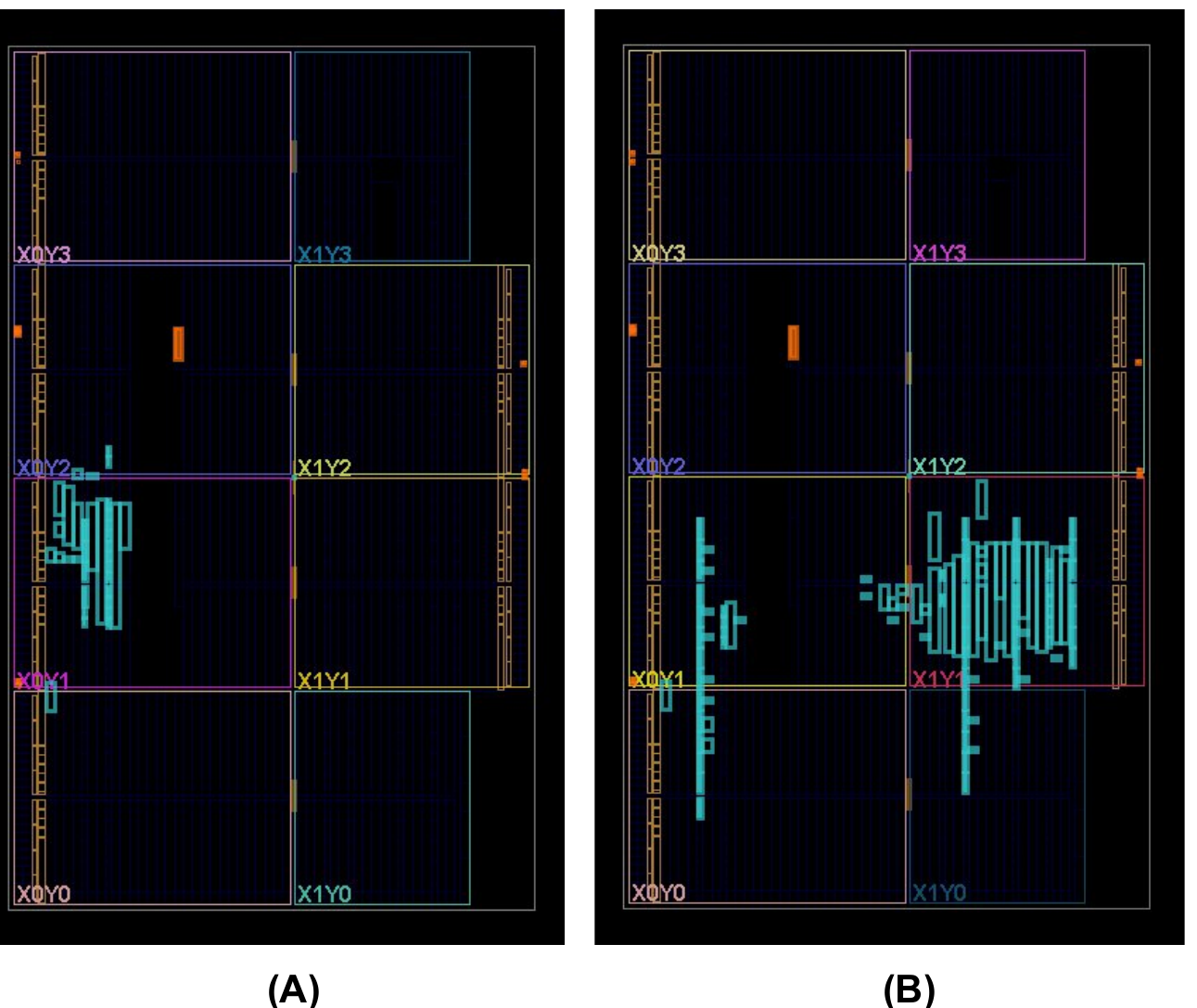}4
\caption{Matched-scale post-route views of the PIS measurement wrappers on XC7A100T: \textbf{(A)} P23 and \textbf{(B)} P129. The larger physical support expands the implemented state/coefficient-memory and routing footprint while the shared arithmetic PE is retained. Quantitative stand-alone-core utilization and timing are given in Table~\ref{tab:resource}.}
\label{fig:board_implementation}
\end{figure}

Two stand-alone cores use the same fixed-point PE but different physical resonator capacities. Table~\ref{tab:resource} shows that reducing the bank from 129 to 23 states lowers LUT use by 65.0\% and BRAM by 88.1\%, while the 14-DSP datapath is unchanged and both designs close the 10-ns constraint. 
Figure~\ref{fig:board_implementation} gives the corresponding implementation-level view. Both panels use the same device scale: P23 retains the shared PE but provisions the smaller state/coefficient bank, whereas P129 carries the larger local memory and routing footprint. The figure is qualitative physical evidence; the resource and timing comparison is reported quantitatively in Table~\ref{tab:resource}. The physical ablation uses three modes to separate runtime activity from instantiated capacity. Three modes are defined:
\begin{itemize}
    \item \textbf{P23/K23:} 23 states are physically provisioned and active.
    \item \textbf{P129/K23:} 129 states are physically provisioned, but the same bitstream schedules only center indices 53-75 (23 states)
    \item \textbf{P129/K129:} the same 129-state bitstream schedules indices 0-128.
\end{itemize}
By setting this up,  P129/K23 versus P129/K129, therefore isolate runtime work while preserving fixed placement and memory. P23/K23 versus P129/K23 keeps the same 23-state logical workload while changing the physically instantiated bank. 
All modes run with a fixed 6,667-cycle frame period. Figure~\ref{fig:results}(b) shows that K23 peaks at 1,316 cycles and K129 at 5,980 cycles; all 100,000 tested frame opportunities meet the deadline. At the same 14,999.25-chirp/s rate, Fig.~\ref{fig:results}(c) gives 42.319$\pm$0.028~mW for P23/K23 and 52.841$\pm$0.027~mW for P129/K23, a 19.912\% reduction from physical right-sizing. P129/K129 measures 52.833$\pm$0.029~mW, with the runtime-only difference from P129/K23 unresolved. Because throughput is matched, the corresponding rail energy is 2.8214~$\mu$J/chirp for P23/K23 versus approximately 3.523~$\mu$J/chirp for the P129 image. The measured result therefore supports the circuit-level distinction between sparse \emph{activity} and sparse \emph{physical capacity}.

\section{Conclusion}

PIS uses an FMCW track before the next observation to create two hardware-relevant forms of sparsity: mean prediction reduces level-crossing activity, while covariance bounds the local event-gap resonator bank. Measured radar, bit-exact RTL, routed Artix-7 cores, physical replay, and repeated VCCINT measurements show that a physically right-sized 23-state implementation reduces logic/memory footprint and rail power, whereas runtime narrowing inside the unchanged 129-state image does not resolve the same benefit. 
PIS remains a tracked mode in which acquisition or reacquisition is retained when the track or local support is invalid, and the measured FPGA power reduction is evidence for this capacity mechanism.
\clearpage
\bibliographystyle{IEEEtran}
\bibliography{refs}
\end{document}